\documentclass[twocolumn,showpacs,preprintnumbers,amsmath,amssymb]{revtex4}

\usepackage{graphicx}
\usepackage{dcolumn}
\usepackage{bm}

\begin{document}

\preprint{APS/123-QED}

\title{Scaling of disordered recursive networks}

\author{Liang Tian}
 \email{tianliang_1985@hotmail.com}
\author{Da-Ning Shi}
 \email{shi@nuaa.edu.cn}
\affiliation{%
College of Science, Nanjing University of Aeronautics and
Astronautics, Nanjing, 210016, PR China
}%

\date{\today}

\begin{abstract}
In this brief report, we present a disordered version of recursive
networks. Depending on the structural parameters $u$ and $v$, the
networks are either fractals with a finite fractal dimension $d_{f}$
or transfinite fractals (transfractal) with a infinite fractal
dimension. The scaling behavior of degree and dimensionality are
studied analytically and by simulations, which are found to be
different from those in ordered recursive networks. The transfractal
dimension $\tilde{d}_f$, which is recently introduced to distinguish
the differences between networks with infinite fractal dimension,
scales as $\tilde{d}_f\sim \frac{1}{u+v-1}$ for transfractal
networks. Interestingly, the fractal dimension for fractal networks
with $u=v$ is found to approach $3$ in large limit of $u$, which is
thought to be the effect of disorder. We also investigate the
diffusion process on this family of networks, and the scaling
behavior of diffusion time is observed numercally as $\tau\sim
N^{(d_{f}+1)/d_{f}}$ for fractal networks and $\tau\sim
\frac{1}{\tilde{d}_f}N$ for transfractal ons. We think that the
later relation will give a further understanding of transfractal
dimension.
\end{abstract}


\maketitle

Complex networks have been studied extensively owing to their
relevance to many real systems such as the world-wide web, the
Internet, energy landscapes and biological and social
networks~\cite{1,2}. Empirical studies indicate that the networks in
various fields exhibit some common topological characteristics: a
logarithmically growing average distance $L$ or diameter $D$ with
network size (small-world property)~\cite{8} and a power-law degree
distribution (scale-free property)~\cite{9}. Recently, Song \emph{et
al} have studied naturally occurring scale-free networks that seem
to be fractal and small-world at the same time~\cite{10}. They
applied a box covering algorithm which enabled them to demonstrate
the existence of self-similarity and fractality in these real
networks. In order to characterize such a behavior, a family of
ordered recursive scale free networks are proposed by Rozenfeld and
ben-Avraham (RA)~\cite{11,12}. By the application of a
renormalization procedure, they exactly solved the fracatily,
dimensionality, and scaling in them. Importantly, a useful measure
of transfinite fractal (transfractal) dimension $\tilde{d}_f$ is
defined and can be used to study the self-similarity and fractality
of the networks with small-world structure (infinite fractal
dimension). However, the determinism in this model prevent it from
describing the diversity and disorder in the nature. In this brief
report, we generalize RA's model and present a disordered version of
recursive networks. Both by theoretical analysis and simulations,
the structure and dimensionality are studied, which are found to be
different from those in previous ordered recursive networks. We also
investigate the diffusion process on this family of networks.

\begin{figure}
\scalebox{0.6}[0.6] {\includegraphics{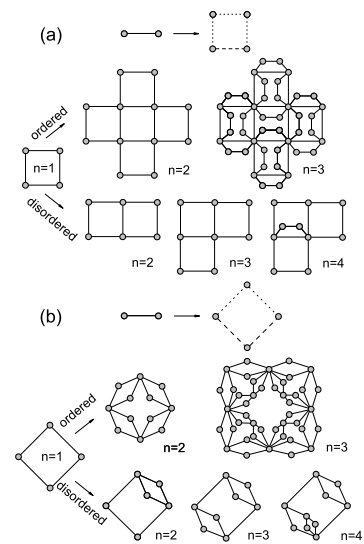}} \caption{\label{f.1}
Illustration of $(u,v)$ flowers and D-flowers. (a) $u =1$ (dashed
line) and $v=3$ (dotted line). (b) $u=2$ and $v=2$. See detailed
method of construction in the main text.}
\end{figure}

Firstly, we give a brief introduction of ordered recursive network
of RA. A special class of their model is $(u,v)$ flowers, where
\emph{every} link in generation $n$ is replaced by two parallel
paths consisting of $u$ and $v$ links ($u\leq v$), to yield
generation $n+1$ (Fig.~\ref{f.1}). A natural choice for the genus at
generation $n=1$ is a ring consisting of $u+v=w$ links and nodes.
The pseudofractal scale-free network (PSFN) proposed by Dorogovtsev,
Goltsev, and Mendes~\cite{13} is just the case of $(1,2)$ flowers.
The disordered version of this model is constructed in the same
iterative manner, except that in each generation not \emph{every}
but only \emph{one random chosen} link is updated. We denote this
family of network by $(u,v)$ disordered flower (D-flower). Examples
of (1, 3), (2, 2) flowers and (1, 3), (2, 2) D-flowers are shown in
Fig.~\ref{f.1}. We can imagine that the (1, 2) D-flower is the same
as the random PSFN proposed by Dorogovtsev, Mendes and
Samukhin~\cite{14}, though it is not displayed here.

\begin{figure}
\scalebox{0.7}[0.7] {\includegraphics{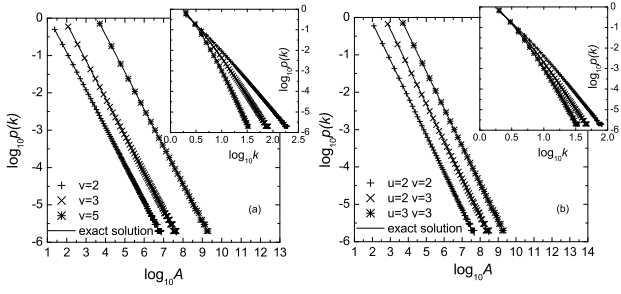}} \caption{\label{f.2}
Log-log plot of degree distribution $p(k)$ for $(u,v)$ D-flowers
versus $A=\prod_{i}^{\omega-1}(k+i)$ for (a) $u=1$ and (b) $u>1$.
The insets show the dependence of $p(k)$ on degree $k$. The
continuous lines correspond to the exact solutions given by
Eq.~(\ref{e.4}). The networks size is $10^6$.}
\end{figure}

According to the recursive algorithm of construction, it is easily
found that the number of links and nodes of a $(u,v)$ D-flower in
generation $n$ is separately $M_n=(\omega-1)n+1$ and
$N_n=(\omega-2)n+2$.

By using the rate-equation approach~\cite{15}, the degree
distribution $p(k)$ can be obtained analytically for $(u,v)$
D-flower. Let $N_n(k)$ be the number of nodes with degree $k$ in
generation $n$, then we can write down the rate equation for the
degree distribution
\begin{equation}\label{e.3}
\frac{dN}{dn}=\frac{(k-1)N_{n}(k-1)-kN_n(k)}{M_n}+\delta_{k,2}.
\end{equation}
In the asymptotic limit $N_n(k)=N_np(k)$ and $M_n=(\omega-1)n$, a
recursive equation is obtained
\begin{eqnarray}\label{e.4}
    p(k)=\left\{\begin{array}{rlc} p(k-1)\frac{k-1}{k+\omega-1},  & \text{for} &    k>2
    \\ \frac{\omega-1}{\omega+1}, & \text{for} &  k=2 \end{array}
    \right.,
\end{eqnarray}
leading to
\begin{equation}\label{e.4}
p(k)=\frac{(\omega-1)\prod_{i=0}^{\omega-1}(2+i)}{\prod_{i=0}^{\omega-1}(k+i)}.
\end{equation}

It should be noted that the degree distribution for disordered
recursive network is different from that ordered one, which is
$p_d(k)\sim k^{-\gamma}$ of degree exponent~\cite{11}
\begin{equation}\label{e.5}
\gamma=1+\frac{\ln(u+v)}{\ln(2)}.
\end{equation}

In Fig.~\ref{f.2}, we report the numerical result of probability
distribution of degree, which is exactly is in agreement with the
theoretical analysis.

In RA's ordered recursive networks, there is a vast difference
between $(u,v)$ flowers with $u=1$ and $u>1$. If $u>1$, the diameter
$D_n$ of the $n$th-generation flower grows as a power of $n$, and
thus it possesses a finite fractal dimension $d_f$. For $u=1$,
however, the diameter scales linearly with $n$, which indicates the
existence of small-world property and the lack of finite fractal
dimension. In order to distinguishes between different networks of
infinite dimensionality, RA define a new measure of dimensionality,
transfractal dimension $\tilde{d}_f$, characterizing how mass scales
with diameter:
\begin{equation}\label{e.6}
D\sim \frac{1}{\tilde{d}_f}\ln N.
\end{equation}

We will demonstrate that these features still hold for disordered
recursive networks. The transfractal dimension $\tilde{d}_f$ for
(1,v) D-flower can be solved by using a mean field
approximation~\cite{16}. We represent each node by the time sequence
it enters the network. Let $d(i,j)$ denotes the distance between
node $i$ and node $j$, and thus the average distance of the model in
the generation $n$ is
\begin{equation}\label{e.6}
L(n)=\frac{2\sigma(n)}{N_n(N_n-1)},
\end{equation}
where the total distance is
\begin{equation}\label{e.7}
\sigma(n)=\sum_{1\leq i<j\leq N_n } d(i,j).
\end{equation}
According to the method of construction for $(1,v)$ D-flower,
$\sigma(n)$ evolves as
\begin{equation}\label{e.8}
\sigma(n+1)=\sigma(n)+S(w-2)+\sum_{i=1}^{N_n}\sum_{j=1}^{\omega-2}d(i,N_n+j),
\end{equation}
where the second term on the right side accounts for the sum of
distance between the new add $w-2$ nodes. In current generation, we
renormalize the updated link  and the two nodes connected to it as a
single node $r$, then
\begin{equation}\label{e.9}
\sigma(n+1)=\sigma(n)+S(w-2)+N_nS(w-2,y)+(w-2)\sum_{i=1}^{N_n}d(i,r),
\end{equation}
where $S(w-2,r)$ denotes the sum of distance between node $r$ and
the new $w-2$ nodes. Since the shortest path from node $r$ to an
arbitrary node $i$ ($i\leq N_n$) never passes the newly added nodes,
we can rewrite $\sum_{i=1}^{N_n}d(i,r)$ as $L(n)(N_n-1)$. Therefore,
in the asymptotic limit $\sigma(n+1)-\sigma(n)=(w-2)d\sigma(N)/dN$,
we obtain
\begin{equation}\label{e.10}
\frac{d\sigma}{dN}=\frac{S(w-2)}{w-2}+\frac{S(w-2,r)}{w-2}N+\frac{2\sigma}{N},
\end{equation}
leading to
\begin{equation}\label{e.11}
\sigma(N)=[\frac{S(w-2,r)}{w-2}\ln N+C]N^2-\frac{S(w-2)}{w-2}N,
\end{equation}
where $C$ is constant independent of $N$. In thermodynamical limit,
incorporating with $L(N)\sim \sigma(N)/N^2$ and $L(N)\sim
\frac{1}{\tilde{d}_f}\ln N$ yields
\begin{equation}\label{e.12}
\tilde{d}_f\sim \frac{w-2}{S(w-2,r)}.
\end{equation}
We note that $S(w-2,r)$ represents the total distance from one to
all the other $w-2$ nodes in a $1$-dimension ring with length $w-1$.
Therefore, it is easily to obtain $S(w-2,r)\sim (w-2)/(w-1)$ for
large $w$, which leads to a scaling behavior of transfractal
dimension of $(1,v)$ D-flower
\begin{equation}\label{e.13}
\tilde{d}_f\sim \frac{1}{w-1}.
\end{equation}

\begin{figure}
\scalebox{0.9}[0.9] {\includegraphics{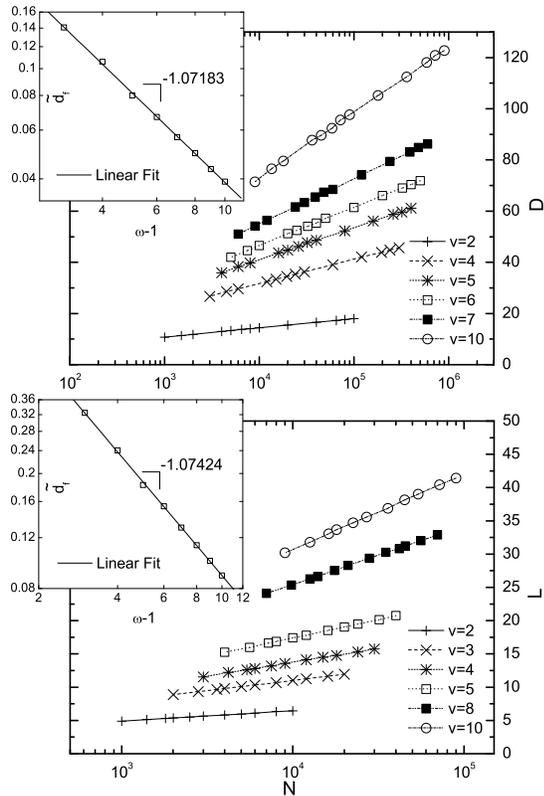}} \caption{\label{f.3} In
$(1,v)$ D-flower, (a) average distance $L$ and (b) diameter $D$
versus network size $N$ for different values of $v$ is showed in
semi-log plot. The linear fit of the data in the main panel
corresponding to the measured transfractal dimension $\tilde{d}_f$
as a function of $w-1$ is displayed in the insets. All the data is
obtained by averaging over $10^3$ different network realizations.}
\end{figure}
\begin{figure}
\scalebox{0.7}[0.7] {\includegraphics{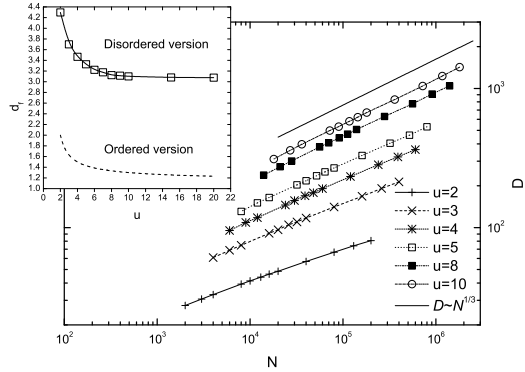}} \caption{\label{f.4} In
$(u,u)$ D-flower with $u>1$, diameter $D$ versus network size $N$
for different values of $u$ is showed in log-log plot. The linear
fit of the data in the main panel corresponding to the measured
fractal dimension $d_f$ as a function of $u$ is displayed in the
inset. For comparison, the exact solution for previous ordered
recursive network of RA (dashed line) is also laid out. All the data
is obtained by averaging over $10^3$ different network
realizations.}
\end{figure}

In Fig~\ref{f.3}, the results for transfractal dimension
$\tilde{d}_f$ obtained by simulations confirms our analytical ones.
In the above derivation, we replace diameter $D$ with average
distance $L$ in the definition of transfractal dimension
$\tilde{d}_f$. Although it brings the quantitative differences, the
scaling behavior of $\tilde{d}_f$ (Eq.~\ref{e.13}) still holds ,
which is demonstrated in the numerical simulations (see
Fig.~\ref{f.3}).

In $(u,v)$ D-flower with $u>1$ possessing a finite fractal dimension
$d_f$, the mean field approximation is invalid, which disable us
from a theoretical analysis. However, we investigate $d_f$ of
D-flower numerically and find some novel properties. We limit it to
a simple case with $u=v$, and the numerical result is displayed in
Fig.~\ref{f.4}. The $d_f$ of $(u,u)$ D-flower decreases
monotonically with $u$, and approach to infinite in the limit of
$u\rightarrow1$, which recovers the previous result. We note that
the quantity of fractal dimension for disordered recursive network
is totally different from that of ordered one~\cite{11}.
Interestingly, it is found that the fractal dimension approaches to
$3$ in the large limit of $u$, while in the ordered case
$d_f\rightarrow1$ in the same limit. We argue that this phenomena is
due the disorder in the construction procedure, which always enlarge
the effective dimension~\cite{17} of the network (see the inset of
Fig.~\ref{f.4}).

\begin{figure}
\scalebox{0.8}[0.8] {\includegraphics{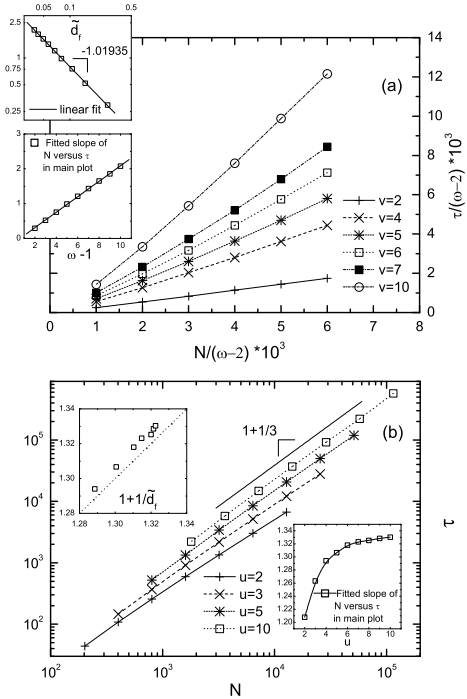}} \caption{\label{f.5}
(a) In $(1,v)$ D-flower, diffusion time $\tau$ versus network size
$N$, which is plotted in rescale manner just for more explicit
sight. The linear fit of the data in main panel as functions of
$\tilde{d}_f$ and $\omega-1$ are shown in the upper and lower insets
separately. (b) $\tau$ versus $N$ for $(u,u)$ D-flower with $u>2$.
The scaling exponent obtained from the linear fitting of the data in
main plot as functions of $1+1/\tilde{d}_f$ and $u$ are displayed in
the upper left and lower right insets separately. The solid line
with slop $1+1/3$ is guide to the asymptotic behavior of the scaling
exponent. All the data is obtained by averaging over $10^4$
independent runs with 100 different configurations of the initial
seed as well as 100 different network realizations.}
\end{figure}

Since we have systematically investigated the dimensionality of
disordered recursive network, one of the most interesting issues is
to check the effects of the fractal dimension $d_f$ or transtractal
dimension $\tilde{d}_f$ on the dynamics taking place upon it. At
present, we briefly consider diffusion of a particle on the $(u,v)$
D-flower. The particle is put on a randomly chosen node of the
network as the seed, and allowed to hop to one of its neighboring
nodes, chosen randomly each time step. In stead of the return
probability~\cite{18}, we consider here the number of nodes visited
by the particle during time $t$, denoted by participation ratio
$P(t)$. Therefore, it takes the values between $P(t=0)=1$ and
$P(t\rightarrow\infty)=N$. To characterize diffusion, we define the
diffusion time $\tau$ associated with the participation ratio by the
condition $P(t=\tau)=cN$ with a constant $c$ between zero and unity.
Numerical simulations of the diffusion are conducted on $(u,v)$
D-flower and the diffusion time $\tau$ is measured. The numerical
factor $c=0.1$ is chosen, and we confirmed other values of $c$ do
not change the scaling behavior of $\tau$.

In Fig.~\ref{f.5}(a), we plot the the diffusion time $\tau$ versus
the network size $N$ of $(1,v)$ D-flower for various values of the
$v$. It is observed that the diffusion time displays a linear
behavior $\tau\sim N$, which is consistence with the infinite
fractal dimension (also infinite effective dimension) of the (1,v)
D-flower and recovers the result on small-world networks~\cite{19}.
Importantly, the linear fitting slope has a scaling relation with
transfractal dimension $\tilde{d}_f$ shown in the upper inset of
Fig.~\ref{f.5}(a), which indicates the behavior
$\tau\sim\frac{1}{\tilde{d}_f}N$. This indicates that, although the
scaling behavior of diffusion time $\tau\sim N$ hold for all of the
networks with infinite fractal dimension, the finer quantitative
difference is distinguished by the transfractal dimension
$\tilde{d}_f$ of those networks, which provides us with a further
understanding of the transfractal dimension. Since
$\tilde{d}_f\sim\frac{1}{\omega-1}$, it is easily obtained
$\tau\sim(\omega-1)N$ displayed in the lower inset of
Fig.~\ref{f.5}(a). The dependence of diffusion time $\tau$ on $N$
for $(u,u)$ D-flower with $u>1$ is shown in Fig.~\ref{f.5}(b). The
scaling behavior $\tau\sim N^{(d_f+1)/d_f}$ is observed (see the
upper left inset of Fig.~\ref{f.5}(b)) in spite of deviation due to
the finite-size effects inherent in numerical calculation of
$d_f$~\cite{20}. From the lower right inset of Fig.~(\ref{f.5}(b)),
we observe that the scaling exponent obtained numerically approaches
to $1+1/3$ as $u$ increases, which is in agreement with the behavior
of fractal dimension that $d_f\rightarrow3$ in large limit of $u$.

In summary, we have investigated the scaling behavior of degree
distribution, dimensionality and diffusion time on a disordered
version recursive networks analytically and numerically. Due to the
introduction of disorder which extensively exists in nature, many
properties emerging in previous ordered recursive networks
dramatically change. Since they possess novel fractality and
dimensionality, it will be interesting to study many physics
possess, such as synchronization and Anderson transition, upon the
networks. Research along this line is in progress.

This work has been partially supported by the Innovation Program for
the Graduate Students of Jiangsu Province of China under grant No.
CX07B-033z and the Foundation for Graduate Students of Nanjing
University of Aeronautics and Astronautics under Grant No.
BCXJ07-11.

\end{document}